\documentclass{jfm}
\usepackage{natbib}
\usepackage{amsmath}
\usepackage{graphicx}

\ifCUPmtlplainloaded \else
  \checkfont{eurm10}
  \iffontfound
    \IfFileExists{upmath.sty}
      {\typeout{^^JFound AMS Euler Roman fonts on the system,
                   using the 'upmath' package.^^J}%
       \usepackage{upmath}}
      {\typeout{^^JFound AMS Euler Roman fonts on the system, but you
                   dont seem to have the}%
       \typeout{'upmath' package installed. JFM.cls can take advantage
                 of these fonts,^^Jif you use 'upmath' package.^^J}%
      }
  \else
  \fi
\fi

\ifCUPmtlplainloaded \else
  \checkfont{msam10}
  \iffontfound
    \IfFileExists{amssymb.sty}
      {\typeout{^^JFound AMS Symbol fonts on the system, using the
                'amssymb' package.^^J}%
       \usepackage{amssymb}%

      }{}
  \fi
\fi

\ifCUPmtlplainloaded \else
  \IfFileExists{amsbsy.sty}
    {\typeout{^^JFound the 'amsbsy' package on the system, using it.^^J}%
     \usepackage{amsbsy}}
    {\providecommand\boldsymbol[1]{\mbox{\boldmath $##1$}}}
\fi

\providecommand\bnabla{\boldsymbol{\nabla}}

\newsavebox{\astrutbox}
\sbox{\astrutbox}{\rule[-5pt]{0pt}{20pt}}

\title[Evaporation instability in a thin film]{Evaporation of a thin film: Diffusion of the vapour and Marangoni instabilities}

\author[E. SULTAN, A. BOUDAOUD, M. BEN AMAR]%
{E\ls R\ls I\ls C\ns S\ls U\ls L\ls T\ls A\ls N%
,\ns
A\ls R\ls E\ls Z\ls K\ls I\ls \ns B\ls O\ls U\ls D\ls A\ls O\ls U\ls D\break
\and M\ls A\ls R\ls T\ls I\ls N\ls E \ns B\ls E\ls N\ns A\ls M\ls A\ls R}

\affiliation{Laboratoire de Physique Statistique, Ecole Normale Sup{\'e}rieure, 24 rue Lhomond, 75231 Paris Cedex 05, France}
\volume{538}
\pagerange{1--50}
\newcommand{\rmn}[1] {\mathrm{#1}}

\newcommand{\mitbf}[1] {\hbox{\mathversion{bold}${#1}$}}
\newcommand{\mc}{\mathcal}
\newcommand{\dd}{{\rm d}}
\newcommand{\kc}{k_\rmn{c}}
\newcommand{\ed}{\,\rmn{e}}
\newcommand{\cc}{\,{\rm c.c}}
\newcommand{\ma}{\mathit{Ma}}
\newcommand{\ca}{\mathit{Ca}}
\newcommand{\mac}{\mathit{Ma}_\rmn{c}}
\newcommand{\cac}{\mathit{Ca}_\rmn{c}}
\newcommand{\pe}{\mathit{Pe}}
\newcommand{\pek}{\mathit{Pe}_\rmn{k}}
\begin{document}
\maketitle
\begin{abstract}
The stability of an evaporating thin liquid film on a solid substrate is investigated within lubrication theory. The heat flux due to evaporation induces thermal gradients; the generated Marangoni stresses are accounted for. Assuming the gas phase at rest, the dynamics of the vapour reduces to diffusion. The boundary condition at the interface couples transfer from the liquid to its vapour and diffusion flux. The evolution of the film is governed by a lubrication equation coupled with the Laplace problem associated with quasi-static diffusion. The linear stability of a flat film is studied in this general framework. The subsequent analysis is restricted to diffusion-limited evaporation for which the gas phase is saturated in vapour in the vicinity of the interface. The stability depends then only on two control parameters, the capillary and Marangoni numbers. The Marangoni effect is destabilising whereas capillarity and evaporation are stabilising processes. The results of the linear stability analysis are compared with the experiments of \cite{christophe} performed in a different geometry. In order to study the resulting patterns, an amplitude equation is obtained through a systematic multiple-scale expansion. The evaporation rate is needed and is computed perturbatively by solving the Laplace problem for the diffusion of vapour. The bifurcation from the flat state is found to be a supercritical transition. Moreover, it appears that the non-local nature of the diffusion problem unusually affects the amplitude equation.

\end{abstract}

\section{Introduction}
\label{intro}

Since the pioneering studies of \cite*{thomson}, \cite*{marg} and \cite*{benard}, much attention has been devoted to the now called Marangoni instabilities. Thomson and Marangoni first proposed surface tension gradients as a cause for convection in liquids. The Marangoni effect consists in the variation of surface tension with temperature or liquid composition and drives this class of instabilities. The hexagonal patterns observed by \cite{benard} in thin layers heated from below prompted a number of studies \cite[for reviews, see][]{davis_rev,schatz}.  Recent research in this field has focused on the correct description of the gas above the fluid layer and of the deformability of the interface \cite[][]{vanhook,golovin}, or the effect of local heating \cite[][]{miladinova,kalliad,yeo}. 

Marangoni effect can also be driven by evaporation. Many experimental situations of interest are reviewed by \cite*{berg}. On the one hand, evaporation generates thermal gradients as the phase transformation requires latent heat. For spreading droplets of slightly volatile liquids, \cite*{redon} reported festoon instabilities near the contact line while \cite{colloid} measured height fluctuations over the whole drop. \cite*{hegseth} observed vigorous interior flow in evaporating droplets of a volatile liquid. On the other hand, in the case of mixtures, a difference of the evaporation rate between components changes the relative concentrations at the interface and so generates surface tension gradients. \cite*{fournier,vuilleumier,fanton,hosoi} studied tears of wine and the associated convection rolls. \cite*{nguyen} showed instabilities induced by surfactants. 
This composition mechanism may enhance evaporation and is therefore useful in drying techniques \cite[][]{marra,obrien,matar}. 

A comprehensive theoretical stability analysis of evaporating/condensing films was done by \cite*{burel}. They included vapour recoil and thermocapillarity, but as in subsequent studies \cite[see][]{oron,margerit,merkt}, the evaporation and condensation are governed by the departure from thermodynamic equilibrium at the interface. Within this framework evaporation is intrinsically destabilising as can be seen in \cite{prosper} who did not consider Marangoni stresses. In the case of very thin films \cite[][]{elbaum} microscopic forces may be destabilising as shown by \cite*{samid,lyush}. It is worth noticing here that all models developed in these papers are one-sided: they do not account for the gas phase dynamics except through the boundary condition at the interface.

In contrast, \cite{deeg_nat,deegan} showed that evaporation of pinned water droplets is limited by diffusion of vapour in air, thus they explained the origin of coffee stains. \cite{cachile} explained their experiments on freely receding evaporating droplets within the same framework. They also observed the drops of certain fluids to loose their axisymmetry and the contact line to become wavy \cite[][]{christophe}. These unexplained instabilities are among the motivations of the present study. In particular, the established one-sided model would always predict an instability for evaporation. 

Our aim here is to generalize the one-sided study of \cite{burel} for evaporating thin films by taking into account the dynamics of the vapour. In section \ref{model}, we build a model which includes both thermodynamically determined transfer of the molecules across the interface and diffusion of vapour in the gas phase. This generalises the two class of models presented above. We describe the liquid film within lubrication theory taking into account surface tension gradients and loss of mass. In section 3, we perform the full linear stability analysis of this system. Then we restrict to the diffusion limited regime which is relevant for the experiments of \cite{christophe}. We find the amplitude equation describing patterns above the instability onset. Eventually, in section 4, we compare our results with the experiments of \cite{christophe}.

\section{The model}
\label{model}
\begin{figure}
\begin{center}
 \includegraphics[width=0.65\textwidth]{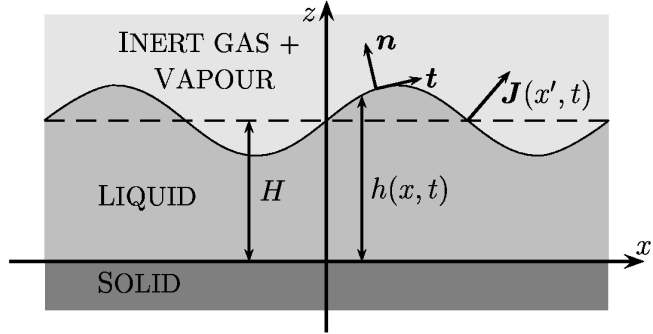}
\end{center}
  \caption{Geometry of the physical system.}
  \label{geom}
\end{figure}

We  consider the dynamics of a  two-dimensional bi-layered liquid-gas system over a solid substrate (figure \ref{geom}). The gas phase is a mixture of an inert gas and of the vapour of the liquid which is volatile. We assume that the gas phase is not saturated by the vapour so that the liquid evaporates. The typical corresponding experimental situation is that of a water layer evaporating in air. The latent heat needed for the phase transformation drives a heat flux to the interface in the liquid. The induced temperature variations may generate surface tension gradients.

The model derived below is built on the lubrication approximation for the liquid layer, and accounts for surface tension variations and loss of mass through evaporation. The gas phase is at rest; its dynamics is reduced to the diffusion of the vapour in the mixture. For a complete description of the system, the boundary conditions at the interface are needed and, in particular, the evaporation rate must be specified. Although this analysis is focused on evaporating films, it applies to the case of condensation as well.

The physical situation of interest is that of volatile liquids such as water and alkanes evaporating in air. A compilation of the physical parameters is given in table \ref{valnum}. The governing equations will be obtained within some approximations, which are relevant to these liquids and will be justified at the end of this section.

We restrict the study to a two dimensional system which coordinates are $x$ along the substrate and $z$ in the normal direction (figure \ref{geom}). The state of the system is determined by the height $h(x,t)$ of the interface and the density $\rho(x,z,t)$ of vapour in the gas phase (both are functions of space and time).  In the following, we write the equations describing the flow and heat diffusion in the liquid phase, diffusion in the gas phase and then the boundary conditions at the interface.

\subsection{The liquid film}
We consider a very thin film so that we work within the long-wavelength approximation (lubrication theory) where the typical height $H$ is much smaller than the typical horizontal scale and we neglect gravity. Our starting point for the  evolution of the thin film is a lubrication equation following \cite*{oron}
\begin{equation}
  \label{lubr}
  \frac{\partial h}{\partial t}+\frac{\partial}{\partial x}\left\{\frac{h^3}{3\mu}\bigg[\gamma\frac{\partial^3h}{\,\partial x^3}+\frac{\partial}{\partial x}\bigg(\frac{J^2}{\rho_\rmn{v}}\bigg)\bigg]+\frac{h^2}{2\mu}\frac{\partial\gamma}{\partial x}\right\}=-\frac{J}{\rho_\ell}
\end{equation}
where $\mu$ is the viscosity of the fluid and $\gamma$ is the surface tension. The last term in the brackets comes from the shear stress due to surface tension gradients. The right-hand side corresponds to the volume loss through evaporation. $\rho_\ell$ is the liquid density, $\rho_\rmn{v}$ the vapour density and $J$ the mass flux across the interface.  Let $J_0$ be the characteristic evaporation rate. The ratio $v_\rmn{ev}=J_0/\rho_\ell$ is the relevant velocity scale in the system. Vapour thrust comes from the back reaction of molecules leaving the liquid into the gas phase; it is a quadratic effect in the evaporation rate. 
Other forces such as Van der Waals are not included. 

\subsection{Surface tension gradient}
\label{surf_tens}
To close equation (\ref{lubr}) we first need to compute the surface tension gradient. 
We use the standard linear equation of state
\begin{equation}
\gamma(T)=\gamma_0-|\displaystyle\frac{\rmn{d}\gamma}{\rmn{d}T}|(T-T_\rmn{subs}).
\label{eq_state}
\end{equation}
where $\gamma_0=\gamma(T_\rmn{subs})$. This approximation is accurate on a large temperature range for most common liquids (not too close from a phase transition). We use the absolute value of $\frac{\dd\gamma}{\dd{T}}(T=T_\rmn{subs})$ as surface tension decreases with temperature for most liquids. The temperature field $T(x,z)$ satisfies the standard convection-diffusion equation in the liquid which reduces to $\frac{\partial^2T}{\partial z^2}=0$ in the long-wavelength approximation. Neglecting density, viscosity, thermal diffusivity and kinetic energy of the gas the energy balance at the interface gives the heat flux \cite[][]{oron}
\[
\kappa\frac{\partial T}{\partial z}(z=h(x))=-{{\mathcal L}_{\rmn{v}}J}
\]
${\mc L}_\rmn{v}$ being the vaporisation latent heat per unit mass, $\kappa$ the thermal conductivity of the liquid. We obtain
\begin{equation}
T(z=h(x))=-\frac{{\mathcal L}_{\rmn{v}}J(x)}{\kappa}h(x)+T_\rmn{subs}
\label{temp}
\end{equation}
where the substrate is assumed isothermal. Equations (\ref{eq_state},\ref{temp}) result in
\begin{equation}
  \label{surft}
  \gamma=\gamma_0+\Big|\frac{\rmn{d}\gamma}{\rmn{d}T}\Big|\frac{{\mathcal L}_{\rmn{v}}}{\kappa}hJ.
\end{equation}
To close the system (\ref{lubr},\ref{surft}) one has to compute the evaporation rate $J$.

It will prove useful in the following to find an upper bound for the amplitude of the temperature variation on the interface. With the help of (\ref{temp}), we obtain: $\Delta T={{\mc L}_\rmn{v}J_0}{\Delta h}/\kappa$, $\Delta h$ being the height fluctuations. The height fluctuations $\Delta h$ are smaller than $H$. Thus, the reduced temperature $\frac{\Delta T}{T_\rmn{subs}}$ is bounded by $\theta=\frac{{\mathcal L}_{\rmn{v}}J_0}{\kappa}\frac{H}{T_\rmn{subs}}$, $H$ being the characteristic thickness of the film. From table 2, we see that $\theta$ is very small for the liquids we are interested in.

\subsection{The vapour}
The gas phase is at rest, so that there is only diffusion of the vapour. We consider the limit of quasi-static diffusion where the characteristic diffusion time is much smaller than the characteristic evaporation time $H^2/D \ll H\rho_\ell /J_0$, $D$ being the diffusion coefficient of the vapour in the gas phase. 
In terms of the P{\'e}clet number $\pe=v_\rmn{ev}\frac{H}{D}$, the latter condition is $\pe\ll1$. Hence, the vapour concentration $\rho(x,z,t)$  (local number of particles per unit volume in the gas phase) is a solution of Laplace{'}s equation:
\begin{equation}
  \label{laplace}
  \bnabla^2\rho =0
\end{equation}
$\bnabla^2=\frac{\partial^2}{\partial x^2}+\frac{\partial^2}{\partial z^2}$ being the 2D-Laplacian.

The gas phase is not saturated by the vapour. 
This condition is enforced by a constant diffusion rate at infinity 
\begin{equation}
\label{bound_infty}
\frac{\partial\rho}{\partial z}\sim-\frac{J_0}{D}, \quad z \to +\infty .
\end{equation}
Experimentally, either the gas is pumped at the top of the container \cite[][]{mancini} or the temperature of a top plate is fixed \cite[][]{vanhook}. In both situations, the gas density is imposed at a certain height above the film, which induces a density gradient. Here, we impose the value of this gradient assuming that the height of the gas phase is much larger than other lengths in the system. 

To solve (\ref{laplace}) a boundary condition at the interface is needed. It is obtained in the next subsection.

\subsection{Evaporation rate}
The vapour and the liquid are coupled through the evaporation rate. The kinetic theory leads to a linear constitutive relation between the mass and the departure from equilibrium at the interface, known as the Hertz-Knudsen relation \cite{prosper}. 
\begin{equation}
  \label{hertz}
  \mitbf{ J}_\rmn{mol}=\alpha\sqrt{\frac{k_\rmn{B}T_\rmn{int}}{2\pi M}}(\rho_\rmn{v}^\rmn{eq}(T_\rmn{int})-\rho|_\rmn{int})\mitbf{ n}
\end{equation}
where 
$M$ is the molecular weight, $\rho_\rmn{v}^\rmn{eq}$ is the density of the gas at the liquid/gas coexistence, $\rho|_\rmn{int}=\rho(z=h(x))$ is the gas density at the interface, $k_\rmn{B}$ is the Boltzmann constant, $\alpha$ is the accommodation coefficient (close to unity) and $\mitbf{ n}$ is the outward normal to the interface. We note $v_\rmn{th}=\alpha\sqrt{\frac{k_\rmn{B}T_\rmn{int}}{2\pi M}}$ which is a typical kinetic velocity.

In the gas phase, the vapour mass flux, related to the departure from uniform vapour density is given by:
\begin{equation}
  \label{fick}
  \mitbf{ J}=-D{\bnabla}\rho.
\end{equation}
Due to the continuity of the normal evaporative flux at the interface, we have 
\begin{equation}
  \label{continuity}
  -D(\mitbf{ n}\cdot\bnabla)\rho|_\rmn{int}=v_\rmn{th}(\rho^\rmn{eq}(T_\rmn{int})-\rho|_\rmn{int}).
\end{equation}
Writing a linear temperature dependant equation of state and using (\ref{temp}), one obtains for the equilibrium density at the interface:
\[
\rho^\rmn{eq}(T)=\rho^\rmn{eq}(T_\rmn{subs})-\frac{\rmn{d}\rho_\rmn{eq}}{\rmn{d}T}\frac{{\mathcal L}_{\rmn{v}}}{\kappa}hJ.
\]
Thus, the boundary condition at the interface (\ref{continuity}) may be rewritten as:
\begin{equation}
  \label{bound_int}
  -D\left(1+v_\rmn{th}\frac{\rmn{d}\rho_\rmn{eq}}{\rmn{d}T}\frac{{\mathcal L}_{\rmn{v}}}{\kappa}h\right)(\mitbf{ n}\cdot\bnabla)\rho|_\rmn{int}=v_\rmn{th}(\rho^\rmn{eq}(T_\rmn{subs})-\rho|_\rmn{int}).
\end{equation}
This is a mixed boundary condition in the sense that it relates the value of the density and its normal gradient at the interface. It corresponds to the conservation of the mass i.e the equality of the evaporation rate and the mass flux in the gas. It includes both diffusion and transfer across the interface which were used separately in the litterature (see subsections 2.6, 2.7 and 4.1).

\subsection{Governing equations. Non dimensional parameters}
\begin{table} 
  \begin{center} 
    \begin{tabular}{l|ccccc} 
      & Water    & Nonane & Octane   & Heptane & Hexane \\ 
      $\rho_\ell$ $\rmn{(kg/m^3)}$  & 1000 & 717 & 699 & 682 & 656  \\
       ${\mc L}_\rmn{v}$ $\rmn{(J/kg)}$ & $2.4\,\,10^{6}$ & $3.18\,\,10^{5}$ & $3.82\,\,10^{5}$ & $3.21\,\,10^{5}$ & $3.22\,\,10^{5}$ \\
      $\gamma$ $\rmn{(kg/s^2)}$ & $7.20\,\,10^{-2}$ & $22.38\,\,10^{-3}$  & $21.77\,\,10^{-3}$& $20.31\,\,10^{-3}$ &  $18.42\,\,10^{-3}$ \\ 
      $\mu$ (kg/m/s) & $8.9\,\,10^{-4}$ & $6.65\,\,10^{-4}$  & $5.08\,\,10^{-4}$ & $3.87\,\,10^{-4}$ & $3.0\,\,10^{-4}$ \\ 
        $\tfrac{\dd\rho^\rmn{eq}}{\dd{T}}$ $\rmn{(kg/m^3/K)}$ & $1.4\,\,10^{-3}$ & $1.7\,\,10^{-3}$ & $5.3\,\,10^{-3}$ & $1.1\,\,10^{-2}$ & $2.7\,\,10^{-2}$ \\ 
      $|\tfrac{\dd\gamma}{\dd{T}}|$ $\rmn{(kg/s^2/K)}$ & $1.5\,\,10^{-4}$ & $9.4\,\,10^{-5}$ & $9.53\,\,10^{-5}$ & $9.80\,\,10^{-5}$ & $1.02\,\,10^{-4}$ \\ 
      $v_\rmn{ev}$ (m/s) & $0.48\,\,10^{-6}$ & $0.2\,\,10^{-6}$ & $0.6\,\,10^{-6}$  & $2.2\,\,10^{-6}$ & $6.8\,\,10^{-6}$\\ 
      $v_\rmn{th}$ (m/s) & $148$ & 52 & $55$  & $58.8$ & $63.3$\\ 
      $\kappa$ (kg.m/s$^3$/K) & $0.60$ & $1.31\,\,10^{-1}$ & $1.28\,\,10^{-1}$ & $1.23\,\,10^{-1}$ & $1.20\,\,10^{-1}$ \\ 
      $D_\rmn{air}$ $(\rmn{m^2/s})$ & $1.9\,\,10^{-5}$ &$3\,\,10^{-6}$& $3\,\,10^{-6}$ &$3\,\,10^{-6}$ &$3\,\,10^{-6}$ \\
    \end{tabular}
    \caption{Values of physical parameters for different fluids at $P=1$ atm and $T=25^\circ$C \cite[][]{handbk}: liquid density $\rho_\ell$, latent heat  $\mathcal{L}_\rmn{v}$, surface tension $\gamma$, liquid viscosity $\mu$, temperature derivative of gas density at equilibrium $\tfrac{\dd\rho^\rmn{eq}}{\dd{T}}$, temperature derivative of surface tension $\tfrac{\dd\gamma}{\dd{T}}$, evaporation velocity $v_\rmn{ev}$, thermal (kinetic) velocity $v_\rmn{th}$, thermal conductivity of the liquid $\kappa$ and diffusion coefficient of the vapour in air $D_\rmn{air}$. $v_\rmn{ev}$ is estimated from the experiments on evaporating drops \cite{christophe} using the formula $v_\rmn{ev}=j_0/R$ where $j_0$ is the evaporation parameter and $R$ is the drop radius; the diffusion coefficient in air $D_\rmn{air}$ is roughly estimated using kinetic theory; using Clapeyron relation, we have $\tfrac{\dd\rho^\rmn{eq}}{\dd{T}}=\tfrac{\rho^\rmn{eq}}{P}\tfrac{{\mc L}_\rmn{v}}{T\Delta v}-\tfrac{\rho^\rmn{eq}}{T}$ where $\Delta v$ is the volume change associated which the vaporizationand $\rho^\rmn{eq}$ is evaluated using the vapour pressure at saturation under 1 atm \cite[][]{handbk}.}
    \label{valnum}
  \end{center} 
\end{table}

For the rescaling of the equations, we choose the typical thickness $H$ of the liquid layer, the characteristic evaporation time $H/v_\rmn{ev}=H\rho_\ell/J_0$, the mass flux far from the substrate $J_0$ and the $J_0H/D$ as units, respectively, of length, time, evaporation rate and density of vapour in the gas phase. We make the substitutions $h\to H\tilde{h}$, $x\to H\tilde{x}$, $z\to H\tilde{z}$, $J\to \rho_\ell v_\rmn{ev}\tilde{J}$, $\rho\to\rho^\rmn{eq}(T_\rmn{subs})+\tilde{\rho}(J_0H/D)$ in equations (\ref{lubr},\ref{surft}) and (\ref{laplace},\ref{bound_infty},\ref{bound_int}). The lubrication equation becomes
\begin{equation}
  \label{lubr2}
  \frac{\partial\tilde{h}}{\partial\tilde{t}}+\ca^{-1}\frac{\partial}{\partial\tilde{x}}\bigg\{\frac{\tilde{h}^3}{3}\frac{\partial}{\partial\tilde{x}}\bigg(\frac{\partial^2{\tilde{h}}}{\,\partial\tilde{x}^2}+\Lambda{\tilde{J}^2}\bigg)+\ma\frac{\tilde{h}^2}{2}\frac{\partial{(\tilde{h}\tilde{J})}}{\partial\tilde{x}}\bigg\}=-\tilde{J}
\end{equation}
where we have introduced $\ca=\frac{\mu v_\rmn{ev}}{\gamma_0}$, $\ma=Hv_\rmn{ev}\rho_\ell\frac{{\mathcal L}_{\rmn{v}}}{\kappa}\frac{1}{\gamma_0}|\frac{\rmn{d}\gamma}{\rmn{d}T}|$ and $\Lambda=\frac{\rho_\ell^2v_\rmn{ev}^2H}{\rho_\rmn{v}\gamma_0}$ respectively the \emph{capillary}, the \emph{Marangoni} and the \emph{vapour thrust} numbers. The Laplace problem reads
\begin{equation}
  \label{lapl_pb2}
  \tilde{\bnabla}^2\tilde{\rho}=\frac{\partial^2\tilde{\rho}}{\partial\tilde{x}^2}+\frac{\partial^2\tilde{\rho}}{\partial\tilde{z}^2}=0
\end{equation}
with the two boundary conditions
\begin{equation}
  \label{nondim_bound}
  (1+\chi\tilde{h})(\mitbf{n}\cdot\tilde{\bnabla})\tilde{\rho}|_\rmn{int}=\pek\tilde{\rho}|_\rmn{int}\quad\mbox{and}\quad\lim_{\tilde{z}\to+\infty}\frac{\partial\tilde{\rho}}{\partial \tilde{z}}=-1.
\end{equation}
where $\pek=\frac{v_\rmn{th}H}{D}$ is a \emph{kinetic P{\'e}clet number} and $\chi=Hv_\rmn{th}\frac{{\mathcal L}_\rmn{v}}{\kappa}\frac{\rmn{d}\rho^\rmn{eq}}{\rmn{d}T}$ is called the thermal expansion number.

The evaporation rate is given by
\begin{equation}
  \label{fick2}
  \tilde{J}=-(\mitbf{n}\cdot\tilde{\bnabla})\tilde{\rho}|_\rmn{int}
\end{equation}
which is the non dimensional version of (\ref{fick}).

\subsection{The one-sided limit}

The evolution equation given by \cite{burel,oron} can be recovered as a particular limit of our model. The case of no diffusion corresponds to the limit of small $\pek$. Introducing a new scaling for the density $\hat{\rho}=\tilde{\rho}\pek/ \chi$, we find that $\hat{\rho}$ satisfies the Laplace equation (\ref{lapl_pb2}) with the boundary conditions
\begin{equation}
  \label{burelb}
  \chi \tilde{J}(1+\chi\tilde{h})=-\hat{\rho}|_\rmn{int}\quad\mbox{and}\quad\lim_{\tilde{z}\to+\infty}\frac{\partial\hat{\rho}}{\partial \tilde{z}}=-\frac{\pek}{\chi}.
\end{equation}
In the limit $\pek\to0$, the boundary condition at infinity yields a uniform density distribution $\hat{\rho}=\rm{Cst}$. The lubrication equation reads
\begin{equation}
  \label{lubr_burel}
  \frac{\partial\tilde{h}}{\partial\tilde{t}}+\frac{1}{\ca}\frac{\partial}{\partial\tilde{x}}\bigg\{\frac{\tilde{h}^3}{3}\frac{\partial}{\partial\tilde{x}}\bigg[\frac{\partial^2{\tilde{h}}}{\,\partial\tilde{x}^2}+\Lambda\bigg(\frac{\chi\hat{\rho}}{1+\chi\tilde{h}}\bigg)^{\!\!2}\bigg]-{\ma}\frac{\tilde{h}^2}{2}\frac{\partial}{\partial\tilde{x}}{\bigg(\frac{\chi\tilde{h}\hat{\rho}}{1+\chi\tilde{h}}\bigg)}\bigg\}=\frac{\chi\hat{\rho}}{1+\chi\tilde{h}}
\end{equation}
where $J$ has been eliminated using (\ref{burelb}). Equation (\ref{lubr_burel}) is the same as the equation obtained by \cite{burel} up to a choice in scalings, when omitting the disjoining pressure term.

\subsection{The pure diffusion limit}
 We now consider the opposite limit $\pek\to\infty$ which we refer to as \emph{diffusion limited regime}. The boundary conditions (\ref{nondim_bound}) for the Laplace problem (\ref{lapl_pb2}) become
\begin{equation}
  \label{nondim_bound2}
  \tilde{\rho}|_\rmn{int}=0\quad\mbox{and}\quad\lim_{\tilde{z}\to+\infty}\frac{\partial\tilde{\rho}}{\partial \tilde{z}}=-1.
\end{equation}

It appears that the gas is saturated in vapour immediately above the interface (in dimensional quantities, $\rho|_\rmn{int}=\rho^\rmn{eq}(T_\rmn{subs})$)  and that evaporation is limited by diffusion. This boundary condition was used in the study of evaporating droplets by \cite{deeg_nat,deegan,christophe,cachile}. The Laplace problem has an electrostatic equivalent: the one of finding the electric potential ($\rho$) with an imposed electric field ($\mitbf{ J}_0$) at infinity and a fixed constant potential on a deformed plane. The sharp edge effect implies a larger evaporation rate at crests which tends to restore the flat state so evaporation is a stabilising mechanism in the diffusion limited regime.

\subsection{Discussion}
The evaluation of the relevant dimensionless parameters for water and different alkanes (table \ref{dimless}) shows that the limits $\pe\to0$, $\theta\to0$ and $\Lambda\to0$ are reasonable. The smallness of the P{\'e}clet number $\mathit{ Pe}$ ensures that the time needed to build up the concentration profile above the film is much smaller than the characteristic time for the motion of the interface, so that stationary diffusion is a good approximation. The smallness of the reduced temperature $\theta$ allows a linear approximation for the gas density at the interface. The relevant physics is therefore contained in the values of the capillary, the Marangoni, the kinetic P{\'e}clet and the thermal expansion numbers $(\ca,\ma,\pek,\chi)$. Other mechanisms, such as molecular interactions, are neglected.


The closed system to be studied consists in the lubrication equation (\ref{lubr2}) coupled to a Laplace problem (\ref{lapl_pb2},\ref{nondim_bound}). This unusual coupling comes from evaporation which relates the film mass loss to the gradient of the vapour concentration. This induces non-locality in the lubrication equation as the mass loss is a function of the whole shape of the interface. To simplify notations, we drop  from now on the tildes for the rescaled variables.
\begin{table}
  \begin{center} 
    \begin{tabular}{c|c|c|ccccc} 
      Number & Definition & Signification & Water & Nonane & Octane   & Heptane & Hexane \\
      $\ca$ $(\times10^8)$ & $\frac{\mu v_\rmn{ev}}{\gamma}$ & $\frac{\mbox{\scriptsize{viscous stresses}}}{\mbox{\scriptsize{capillary stresses}}}$ & $0.6$ & $0.6$ & $1.4$ & $4.2$ & $11$ \\
      $\ma$  $(\times10^6)$ & $H\frac{{\mathcal L}_{\rmn{v}}v_\rmn{ev}\rho_\ell}{\kappa\gamma}|\frac{\rmn{d}\gamma}{\rmn{d}T}|$ & $\frac{\mbox{\scriptsize{Marangoni stresses}}}{\mbox{\scriptsize{capillary stresses}}}$ & $0.7$  & $0.3$ & $1.1$ & $3.8$ & $13$\\ 
      $\pe$ $(\times10^9)$ & $v_\rmn{ev}\frac{H}{D}$ & $\frac{\mbox{\scriptsize{evaporation time}}}{\mbox{\scriptsize{diffusion time}}}$ &$4.9$ & $2.0$ & $6.1$  & $22$ & $69$ \\ 
       $\pek$ & $v_\rmn{th}\frac{H}{D}$ & $\frac{\mbox{\scriptsize{kinetic time}}}{\mbox{\scriptsize{diffusion time}}}$ & 1.5 & 3.5 & 3.7  & 3.9 & 4.2 \\ 
       $\chi$ & $v_\rmn{th}\frac{\rmn{d}\rho^\rmn{eq}}{\rmn{d}T}\frac{{\mathcal L}_{\rmn{v}}H}{\kappa}$ & \scriptsize{density fluctuations} & 0.16 & 0.05 & 0.20 & 0.38 & 1.03 \\ 
       $\frac{\pek}{\chi}$ & $\frac{\kappa}{D({\rmn{d}\rho^\rmn{eq}}/{\rmn{d}T})\mathcal{L}_\rmn{v}}$ & 
$\begin{array}{c}
  \mbox{\scriptsize{diffusion limited v.s.}} \\
  \mbox{\scriptsize{one-sided regime}}
\end{array}$
 & 9.4 & 70 & 18.5  & 10.3 & 4.1 \\ 
      $\theta$ $(\times10^{6})$ & $\rho_\ell v_\rmn{ev}\frac{{\mathcal L}_{\rmn{v}}}{\kappa}\frac{H}{T_\rmn{subs}}$ & \scriptsize{temperature fluctuations}  &$1.3$  &$0.4$ & $1.1$ & $4.0$ & $12$ \\ 
    $\Lambda$ $(\times10^{11})$ & $\frac{\rho_\ell^2v_{\rmn{ev}}^2H}{\rho_\rmn{v}\gamma_0}$ & $\frac{\mbox{\scriptsize{vapour thrust}}}{\mbox{\scriptsize{capillarity}}}$  &$2.8$  &$0.6$ & $1.9$ & $9.0$ & $31$ \\   
    \end{tabular}
    \caption{Typical values for the non-dimensional parameters.}
    \label{dimless}
  \end{center}
\end{table}

\section{Stability of the flat interface}
\label{stabty}
Equations (\ref{lubr2},\ref{lapl_pb2}) have as solution for the film thickness $h(x,t)=1-t$ and gas density $\rho(x,z,t)=-z+C-(1-\chi/ \pek)t$ with $C=(\pek-1-\chi)/\pek$. As this base state is non stationary, linearisation of the equations gives  a non autonomous partial differential equation, so that standard linear stability (modal) analysis should not apply.
For simplicity, we  assume from now on that the base state is $h(x,t)=1$ and $\rho(x,z,t)=-z+C$, which amounts to adding a volume source $v_\rmn{ev}$ in the right-hand side of equation (\ref{lubr}), as this source compensates exactly for the loss of mass at infinity. This also amounts to a quasi-steady approximation for which evaporation is sufficiently slow so that the thickness of the layer remains approximately constant during the growth of unstable modes. 

\subsection{Full linear stability analysis}
We study the stability of the flat state by seeking solutions of equations (\ref{lubr2},\ref{fick2}) in the form:
\[
h=1+\delta h,\quad\rho=C-z+\delta\rho,\quad J=1+\delta J.
\]
After linearisation, those equations admit Fourier-mode solutions of wavenumber $k$ and growth rate $\Omega$: $\delta h=A\ed^{\Omega t}\ed^{ikx}+\cc$ ($\cc$ stands for the complex conjugate of the preceding term); we now compute the corresponding $\delta\rho$ and $\delta J$. To begin with, as a harmonic function, $\delta\rho$ has to be of the form $B\ed^{ikx}\ed^{-|k|z}+\cc$ ($B$ is a complex valued function of $k$), where the absolute value must be taken in order to ensure vanishing $\delta\rho$ at $z\to+\infty$.
The boundary condition (\ref{nondim_bound}) at the interface gives at linear order $\delta\rho|_{z=1}=\frac{\pek-\chi}{|k|(\chi+1)+\pek}\delta h$. Hence, we can compute $B(k)$
\begin{equation*}
  \delta\rho=\frac{\pek-\chi}{|k|(\chi+1)+\pek}\delta h\ed^{-|k|(z-1)}+\cc.
\end{equation*}
From (\ref{fick2}) we get the perturbed evaporation rate $\delta J=-\frac{\partial}{\partial z}\delta\rho|_{z=1}$. Plugging in (\ref{lubr2}) and dropping non-linear contributions gives the dispersion relation:
\begin{equation}
  \label{omega1}
  \Omega(k)=-\frac{1}{3\ca}k^4+\frac{\ma}{2\ca}k^2-|k|\frac{\pek-\chi}{|k|(\chi+1)+\pek}.
\end{equation} 
$\Omega(k)$ is growth rate of the wavenumber $k$; if $\Omega(k)>0$, then the perturbation grows and the corresponding mode is unstable. Evaporation can either stabilise or destabilise large wavelengths depending on the values of $\pek$ and $\chi$ whereas capillarity stabilises short wavelengths; Marangoni effect drives the instability.

\subsection{Transfer rate limited regime: linear stability analysis}
In the limit $\pek\ll\min\{\chi,k,k\chi\}$, corresponding to the one-sided model, the dispersion relation (\ref{omega1}) becomes
\begin{equation}
  \label{omega_pek}
  \Omega(k)=-\frac{1}{3\ca}k^4+\frac{\ma}{2\ca}k^2+\frac{\chi}{1+\chi}.
\end{equation}
meaning that the film is always unstable. This limit was studied in detail by  by \cite{burel}. In fact, it is a singular limit (the base state has a different form), and spurious effects arise at $k=0$ : equation (\ref{omega_pek}) predicts exponential growth of a constant change in the film thickness. Accounting for the time dependance of the base state as done by \cite{burel} corrects this artifact (it yields in particular $\Omega(0)=0$) but does not change the unstable behaviour.

\subsection{Diffusion limited regime: linear stability analysis}

\begin{figure}
  \begin{center}
    \includegraphics[width=0.65\textwidth]{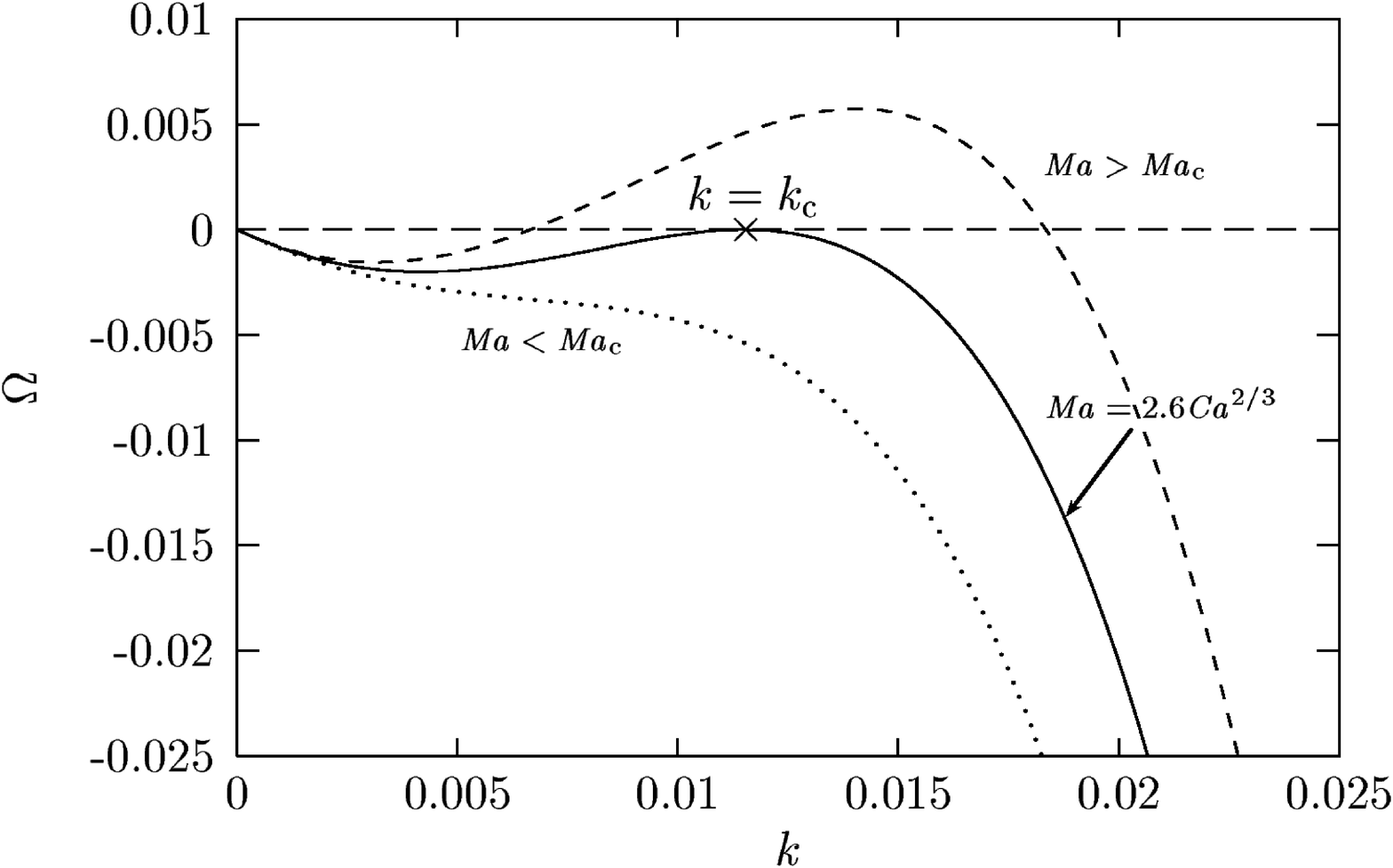}
  \end{center}
  \caption{Diffusion-limited regime: growth rate $\Omega$ of the mode of wave number $k$ (dimensionless quantities) for three typical value of the Marangoni number $\ma$.  We have chosen here a capillary number $\ca=10^{-6}$.}
  \label{disp_rel}
\end{figure}

We now consider the opposite limit $\pek\gg\max\{\chi,k\}$ which is reasonable given the experimental parameters. 
In this regime the dispersion relation (\ref{omega1}) becomes
\begin{equation}
  \label{omega}
  \Omega(k)=-\frac{1}{3\ca}k^4+\frac{\ma}{2\ca}k^2-|k|.
\end{equation}
The last term shows again that evaporation stabilises long wavelengths in this regime (figure \ref{disp_rel}). The absolute value of the wavenumber $k$ comes from quasi-static diffusion. Such a non-analyticity in the dispersion relation is well-known in the context of diffusive growth \cite[see e.g.][]{langer}.
The relevant control parameters are the capillary and Marangoni numbers. Suppose that $\ca$ is fixed; when the Marangoni number $\ma$ is small, there is no unstable mode. The first unstable wave number $\kc$ (marginally stable mode) appears when the Marangoni number reaches a critical value $\mac$ such that $\Omega(k=\kc)=0$ and $\frac{\dd}{\dd{k}}\Omega(k=\kc)=0$. Solving this system, we obtain 
\begin{equation}
  \label{margin}
  \left\{
    \begin{array}{l}
      \mac=18^{1/3}\ca^{2/3}\medskip\\
      k_\rmn{c}^2={\displaystyle\frac{\mac}2}
    \end{array} 
  \right.
\end{equation}
for the critical parameters at the threshold.

The relevance of these results to the experiments is examined in the last section.

\subsection{Diffusion limited regime: weakly non-linear analysis}
The preceding linear stability analysis shows that the system may become unstable, it gives the most unstable wavelength and the instability threshold. However, predicting the nature of the transition or the observed patterns requires a more refined treatment. For instance, if the transition is discontinuous, the pattern might be very different from the linearly most unstable mode. This is why we perform a weakly nonlinear analysis close to the critical point. It is restricted to the diffusion limited regime (equations \ref{lubr2} and \ref{lapl_pb2},\ref{nondim_bound2}).

In this analysis, nonlinear contributions to the evaporation rate are needed; they are computed in appendix \ref{v_sheet}:
\begin{eqnarray}
  \label{evap_expn}
  \begin{array}{lll}
   J[h]&=&1-\frac{\partial}{\partial{x}}{\mc H}[h]+\left\{\frac{1}{2}\left(\frac{\partial{h}}{\partial{x}}\right)^2+\frac{\partial^2h}{\partial{x}^2}h+\frac{\partial}{\partial{x}}{\mc H}\left(h\frac{\partial}{\partial{x}}{\mc H}[h]\right)\right\}\\
&\phantom{=\,}&+\left\{\frac{1}{2}\left(\frac{\partial{h}}{\partial{x}}\right)^2{\mc H}\big[\frac{\partial h}{\partial{x}}\big]-\frac{1}{2}\frac{\partial^2}{\partial{x}^2}\left(h^2\frac{\partial}{\partial{x}}{\mc H}[h]\right)-2h\frac{\partial{h}}{\partial{x}}{\mc H}\big[\frac{\partial^2h}{\partial{x}^2}\big]\right.\\
&&\left.\phantom{\left(\frac{h}{x}\frac{h}{x}\frac{h}{x}\right)^2}-\frac{\partial}{\partial{x}}{\mc H}\left(h\frac{\partial}{\partial{x}}{\mc H}\left(h\frac{\partial}{\partial{x}}{\mc H}[h]\right)\right)-\frac{1}{2}\frac{\partial}{\partial{x}}{\mc H}\big[h^2\frac{\partial^2{h}}{\partial{x}^2}\big]\right\}\\
&\phantom{=\,}&+ {\mc O}(h^4).
\end{array}
\end{eqnarray}
where ${\mc H}$ is the Hilbert transform (appendix \ref{hbt}). Each successive correction to the base state is an integro-differential transform of the interface profile $h$; this comes from the non-local nature of the Laplace problem.

We use a multi-scale expansion which is valid when the spatial Fourier spectrum of $h(x,t)$ is concentrated around $\kc$ \cite[see e.g][]{mville}. We look for an equation of evolution for the slowly varying function $A(X,T)$ such that $h(x,t)=A(X,T)\exp(i\kc x)+\cc$. Formally, we use $\epsilon$ as an expansion parameter. We assume that $h$ is a function of both the fast scales $x,t$ and the slow scales $X=\epsilon x,\ T=\epsilon^2t$. This choice for the slow scales is the natural one given that $\Omega(k)$ is maximum at $\kc$. We consider the neighbourhood of the marginal stability and we rescale the control parameter as
\[
\Omega(\kc)=\epsilon^2\omega(\kc).
\]
From the chain rule for differentiation, we make the replacements
\[
\partial_x \to \partial_x+\epsilon\partial_X,\quad \partial_t \to \partial_t +\epsilon^2 \partial_T.
\]
We also assume that $h$ can be expanded as
\[
h(x,t)=1+\epsilon h_1(x,t)+\epsilon^2 h_2(x,t)+\cdots
\]

The procedure to obtain the amplitude equation (equation for $A(X,T)$) is quite standard and is detailed in appendix \ref{wnl}. However the present case has the peculiarity that the expansion must be pursued up to order 6 as $h_1$ is found to vanish. This is due to the coupling between the evaporation rate and the $k=0$ eigenmodes of the linearised evolution operator.

Taking the rescalings $X\to X/k_\rmn{c}$ and $T \to T/k_\rmn{c}$, neglecting terms of order $\kc^2$ and higher which is consistent with lubrication theory, the amplitude equation reads
\begin{align}
\frac{\partial A}{\partial T}=&\sigma A+\frac{3}{2}\frac{\partial^2A}{\,\partial X^2}-2i\frac{\partial^3A}{\,\partial X^3}-\frac{1}{2}\frac{\partial^4A}{\,\partial X^4}-\frac{45}8|A|^2A\nonumber
\\
&+\displaystyle\frac{3}{2}\kc A\left\{ {\mc H}\int\!\!\!\int\left(\left\{\sigma-\frac{\partial}{\partial T}\right\}{\mc H}+\frac{\partial}{\partial X}\right)|A|^2+2i\int{\mc H}\left[A\frac{\partial\overline{A}}{\partial X}\right]-4i{\mc H}[|A|^2]\right\}
\label{amplt}
\end{align}
the control parameter being
\begin{equation}
  \sigma=2\frac{\ca-\cac}{\cac}+\frac{3}2\frac{\ma-\mac}{\mac}.
  \label{ctrl_prm}
\end{equation}
As far as we are aware of, this non-local kind of amplitude equation has not been derived before. The non-local terms are important for large amplitude patterns or for finite size systems. 

The solution $A=0$ of the amplitude equation becomes unstable when $\sigma>0$ (which is consistent with the linear stability). When $\sigma>0$, there are stationary solutions of the form $A(X,T)=A_0 \exp(iQX)$ with $\sigma=45/8\,|A_0|^2+3/2\,Q^2+2\,Q^3+1/2\,Q^4$. They correspond to a stationary pattern with thickness fluctuation $h(x,t)=A_0 \exp\{i(\kc+Q)x\}$, which is modulated around the critical wavenumber $\kc$. 
Thus, as  the prefactor of $|A|^2A$ is negative, the transition from the flat state ($A=0$) to a state with height fluctuations is a supercritical (continuous) pitchfork bifurcation in contrast with the studies of \cite{vanhook} or \cite{thiele} on heated fluid layers.


\section{Discussion}
\label{discn}

\subsection{Transfer rate limited versus diffusion limited evaporation}
We obtained (\ref{continuity}) as a boundary condition prescribed at the 
interface combining both the transfer rate across the interface and the 
diffusion of the vapour in the gas phase. This boundary condition might be 
simplified in two distinct limits according to the values of the kinetic 
P{\'e}clet number and the thermal expansion number. These two limits have 
been used separately in the literature; however, it has been overlooked 
that they fall within a general framework, although \cite{margerit} have 
treated vapour diffusion in the case of no evaporation. 

In the first limit, $\pek\ll\min\{\chi,k,k\chi\}$, the evaporation process 
is limited by the transfer of molecules across the interface. The 
diffusion of the vapour can be ignored and the classical Hertz-Knudsen 
(\ref{hertz}) relation gives the evaporation rate. It is worth noticing 
here that transfer-limited evaporation is always destabilising, if 
the free surface undergoes a small shape perturbation, liquid portions 
that are closer to (resp. farther from) the substrate evaporate faster 
(resp. slower) and the disturbance is amplified. The so-called one-sided 
models for evaporating layers 
\cite[][]{prosper,burel,samid,lyush,margerit,merkt} correspond to this 
limit as they do not consider the dynamics of the gas phase and so discard 
vapour diffusion.

In the second limit, $\pek\gg\max\{\chi,k\}$, the evaporation process is 
limited by the diffusion of the vapour in the gas phase. The gas phase is 
saturated in vapour immediately above the interface and the evaporation 
flux is given by the Fick relation (\ref{fick}). The electrostatic 
analogue of the Laplace problem for the vapour density, and the related sharp edge effect show that the evaporation rate is larger at bumps hence diffusion 
limited evaporation is stabilising. The study of evaporating droplets by 
\cite{deegan} and \cite*{cachile} fall within this approximation. 

It appears that this second limit is of large validity. Indeed, the ratio 
$\pek/\chi$ which depends only on the nature of the fluid is the range 
4--70 for the fluids considered here, whereas $\pek>1$ (a more restrictive 
condition) as soon as the thickness is larger than $D/v_{\mathrm{th}}\sim 
0.1 \ \mathrm{\mu m}$, which is the thickness scale determined by the 
diffusivity of vapour $D$ and the thermal velocity $v_{\mathrm{th}}$. 

\subsection{Comparison with experiments on evaporating droplets}
We turn now to the experimental relevance of our analysis. Despite the 
number of theoretical studies on evaporating thin films, very few 
experiments have been conducted. \cite{redon}, \cite{colloid} and 
\cite{christophe} observed instabilities in the shape of evaporating 
droplets. In the first set of experiments \cite[][]{colloid,redon}, the 
spreading of drops of silicon oils was studied. As silicon oils have a 
very low volatility, evaporation rate is small and the shear stress 
associated with the spreading is important. In the second set of 
experiments \cite[][]{christophe} evaporation sets the velocity scale. 
This is why we focus on this latter set.

\cite{christophe} studied receding evaporating drops of water and alkanes 
on smooth substrates. The data in tables 1 and 2 corresponds to this 
experimental situation. For water, heptane and hexane the drop looses its 
axisymmetry and develops a regular wavy pattern near the contact line : 
the height fluctuates with a well-defined wavenumber.
 No instabilities are observed neither with octane nor nonane. We now 
compare with the stability analysis in the diffusion limited regime 
(\ref{margin}).
This comparison requires the choice of a thickness lengthscale; we retain 
the typical thickness of the unstable region, $h\sim200\mbox{ }\rm{nm}$. 
We estimate the typical evaporation rate in this zone using 
$J(r)=J_0/\sqrt{1-(r/R)^2}$ \cite[][]{deegan}, $R$ is the drop radius and 
$r$ is the distance to the drop axis. In the experiments, the radius $R$ 
is of order $0.5\mbox{ }\rmn{mm}$ and the size of the unstable zone is 
$R-r\sim5\mbox{ $\mu$m}$.
\begin{table} 
  \begin{center} 
    \begin{tabular}{l|cccc} 
      & $\ca$   & $\ma$  & $\mac(\ca)$ & $\mac/\ma$\\ 
      Water & $5.90\,\,10^{-9}$ & $6.66\,\,10^{-7}$ & $8.59\,\,10^{-6}$ & 
$12.9$ \\
      Nonane & $5.94\,\,10^{-9}$ & $2.89\,\,10^{-7}$ & $8.60\,\,10^{-6}$ & 
$30$ \\
      Octane & $1.4\,\,10^{-8}$ & $1.10\,\,10^{-6}$ & $1.52\,\,10^{-5}$ & 
$13.8$ \\      Heptane & $4.19\,\,10^{-8}$ & $3.76,\,10^{-6} $ & 
$3.16\,\,10^{-5}$ & $8.4$ \\
      Hexane & $1.11\,\,10^{-7}$ & $1.33\,\,10^{-5}$ & $6.05\,\,10^{-5}$ & 
$4.55$ \\ 
    \end{tabular}
    \caption{Values of the control parameters $(\ca,\ma)$ for different 
volatile fluids and comparison with the instability threshold in the diffusion limited regime.}
    \label{fluids_prms}
  \end{center} 
\end{table}
According to table \ref{fluids_prms} droplets of all fluids are stable. 
However, the geometry of a droplet is different from that of a constant 
thickness film analysed in this paper so we cannot conclude directly on 
the stability. We should only compare the relative values of the ratio 
$\mac/\ma$ between different fluids. This ratio is of order one meaning 
that the distance to the threshold is small. Moreover, the fluids can be 
sorted from the less stable to the most stable as: hexane, heptane, water, 
octane and nonane. This compares well with the experiments: festoon 
patterns have been observed with hexane, heptane and water and none with 
octane and nonane. However, caution should be taken with water as 
evaporating droplets of water have an anomalous retraction velocity 
\cite[][]{christophe}. Equation (\ref{margin}) gives as wavelengths 
$\lambda_\rmn{Heptane}\simeq320\mbox{ $\mu$m}$ and 
$\lambda_\rmn{Water}\simeq600\mbox{ $\mu$m}$ which are an order of 
magnitude larger than the experimental 
$\lambda_\rmn{Heptane}\simeq50\mbox{ $\mu$m}$ and 
$\lambda_\rmn{Water}\simeq30\mbox{ $\mu$m}$. To summarize, the comparison 
with experiments is satisfying as we predict an instability threshold in 
contrast with the one-sided model (see section 3.2), and we find that the 
stability increases with the weight of the alkane.

\subsection{Main results}
In this paper, we constructed a two-sided model for evaporating thin 
liquid films. In order to predict quantitatively the evaporation rate, we 
have considered both transfer rate across the interface and diffusion of 
the vapour in the gas phase. The experiments of \cite{christophe} 
motivated the study of the regime for which evaporation is 
diffusion-limited. In this context, the system describing the evolution of 
the height profile $h$ couples the lubrication of the substrate by the 
thin film to the diffusion of its vapour in the gas phase. Using a linear 
stability analysis, we predicted an instability threshold and classified 
the stability for different liquids, in agreement with the experiments.
The results of the linear stability analysis are not too far from \cite{christophe} experiments on evaporating droplets. To push the comparison further, it would be interesting to perform experiments on extended flat films. 

Moreover, the diffusion equation confers a non-local aspect to the 
dynamics of the film which is found to persist in the amplitude equation 
(\ref{amplt}) established  within a weakly nonlinear study. This property 
was unexpected and contrasts with \cite*{engel} results on the Rosensweig 
instability of magnetic fluids; it originates from the presence of uniform 
profiles in the null-space of the linearised evolution operator (see 
appendix \ref{wnl}).

This study suggests investigation of more complicated situations such as 
when both transfer rate and diffusion of the vapour are important (i.e. 
when $\pek$ has a finite value), in the non-linear regime. As analytical 
extensions to the present study seem difficult, numerical computations of 
the complete two-sided model would allow the investigation of effects such 
as the non-stationary of the base state, the finite extension of the system, 
three dimensional patterns or the geometry of a droplet. Consequences of 
the non-local terms should also be interesting to further study.

\begin{acknowledgements}
We are very grateful to Anne-Marie Cazabat and Christophe Poulard for 
getting us intersted in the stability of evaporating droplets.
\end{acknowledgements}
\appendix
\section{The evaporative flux. Electrostatic analogy.}
\label{v_sheet}
This appendix is devoted to the perturbative treatment of the Laplace problem associated with the diffusion limited regime
\begin{equation*}
  \left\{\begin{array}{l}
      \bnabla^2\rho=0\smallskip\qquad(z>h(x))\\
      \rho(x,h(x))=0 \smallskip\\
      \frac{\partial\rho}{\partial z}(x,z=+\infty)=-1
    \end{array}\right.
\end{equation*}
where $h=h(x)$ is a given regular (bounded and differentiable) function. Precisely, the point is to compute $-(\mitbf{n}\cdot\bnabla)\rho|_{z=h(x)^+}$. 
\subsection{Vortex sheet formalism}
Using electrostatic terminology, the problem is to find the electric field $\mitbf{J}$ immediately above a deformed charged plane. Introducing a superficial charge distribution $\rho=\sigma(x)\delta(z-h(x))$ ($\delta(z-a)$ is a Dirac mass concentrated at the point $z=a$), one can write the integral representation:
\begin{equation}
  \label{int_rep}
  \mitbf{J}(x,z)=\int\sigma(x')\frac{(x-x')\mitbf{e}_x+(z-h(x'))\mitbf{e}_z}{(x-x')^2+(z-h(x'))^2}\dd{\ell(x')}\qquad(z\neq h(x))
\end{equation}
$\dd{\ell}=\sqrt{1+(\frac{\dd h}{\dd x})^2}\,\dd x$ being the arc-length element.

At the interface, two boundary conditions are prescribed. To begin with, according to Gauss theorem, the field has a normal discontinuity
\begin{equation}
  \label{discty}
  (\mitbf{J}^+-\mitbf{J}^-)\cdot\mitbf{n}=2\pi\sigma\frac{\dd\ell}{\dd{x}}
\end{equation}
where $\mitbf{J}^+$ (resp. $\mitbf{J}^-$) stands for the field just above (resp. below) the interface $z=h(x)$. Moreover, the tangential component has to vanish in order to fulfil the condition $\rho(x,h(x))=0$:
 \begin{equation}
   \label{cty}
  \mitbf{J}^+\cdot\mitbf{t}=\mitbf{J}^-\cdot\mitbf{t}=0
\end{equation}
Now, setting $z=h(x)$ in (\ref{int_rep}), one has to take the Cauchy principal value (denoted as $\mathrm{PV}$) of the integral in order to get a well defined expression. This regularised integral is equal to the half-sum of $\mitbf{J}^+$ and $\mitbf{J}^-$ so that the conditions (\ref{discty},\ref{cty}) may be rewritten as:
\begin{equation}
  \label{above}
  J^+=\pi\sigma\sqrt{1+\left(\frac{\dd{h}}{\dd{x}}\right)^2}+\frac{1/4}{\sqrt{1+\left(\frac{\dd{h}}{\dd{x}}\right)^2}}{\rm PV}\int\frac{\sigma(x')}{x-x'}\frac{-\frac{\dd{h}}{\dd{x}}(x)+\phi(x,x')}{1+\phi(x,x')^2}\dd{\ell(x')}
\end{equation}
and
\begin{equation}
\label{tang}
{\rm PV}\int\frac{\sigma(x')}{x-x'}\frac{1+\frac{\dd{h}}{\dd{x}}(x)\phi(x,x')}{1+\phi(x,x')^2}\dd{\ell(x')}=0
\end{equation}
with $\phi(x,x')=\frac{h(x)-h(x')}{x-x'}$ and $J^+= \mitbf{J}^+\cdot\mitbf{n}=\|\mitbf{J}^+\|$.
\subsection{Perturbative treatment}
The general relation (\ref{tang}) implicitly gives the superficial charge $\sigma$ as a function of $h$. Since the inversion is not possible analytically, we assume that the deflection from the flat plane $h$ is weak; we introduce a small parameter $\eta$ such as $h$ is replaced by $\eta h$ (a possible choice is to set $\eta=\sup_x|h(x)|$) and write a perturbative expansion $\sigma=\sigma^{(0)}+\eta\sigma^{(1)}+\eta^2\sigma^{(2)}+\ldots$. The $\eta=0$ contribution corresponds to the plane interface for which the field is uniform; the boundary condition $\mitbf{J}(x,z=+\infty)=J^{(0)}\mitbf{e}_z$ readily gives $\sigma^{(0)}=J^{(0)}/\pi$. Solving (\ref{tang}) up to ${\mc O}(\eta^4)$, we find
\begin{equation}
\label{surfdist}
\sigma[h]=\frac{J^{(0)}}{\pi}\bigg(1+\eta^2\bigg\{\frac{1}{2}\left(\frac{\dd{h}}{\dd{x}}\right)^2+h\frac{\dd^2h}{\dd{x}^2}+\frac{\dd}{\dd{x}}{\mc H}\bigg[h\frac{\dd}{\dd{x}}{\mc H}[h]\bigg]\bigg\}+ {\mc O}(\eta^4)\bigg)
\end{equation}
where $\mc {H}$ is the Hilbert transform (see Appendix \ref{hbt}).

The remaining step is simply to plug (\ref{surfdist}) into (\ref{above}). Without having to compute further terms for $\sigma(x)$, we can express the evaporative flux expansion up to ${\mc O}(h^4)$. Setting $\eta=1$, we find
\begin{eqnarray*}
  J[h]&=&1-\displaystyle\frac{\dd}{\dd{x}}{\mc H}[h]+\left\{\frac{1}{2}\left(\frac{\dd{h}}{\dd{x}}\right)^2+\frac{\dd^2h}{\dd{x}^2}h+\frac{\dd}{\dd{x}}{\mc H}\left(h\frac{\dd}{\dd{x}}{\mc H}[h]\right)\right\}\\
&\phantom{=\,}&+\left\{\displaystyle\frac{1}{2}\left(\frac{\dd{h}}{\dd{x}}\right)^2{\mc H}\bigg[\frac{\dd h}{\dd{x}}\bigg]-\frac{1}{2}\frac{\dd^2}{\dd{x}^2}\left(h^2\frac{\dd}{\dd{x}}{\mc H}[h]\right)-2h\frac{\dd{h}}{\dd{x}}{\mc H}\bigg[\frac{\dd^2h}{\dd{x}^2}\bigg]\right.\\
&&\left.\phantom{\left(\displaystyle\frac{h}{x}\frac{h}{x}\frac{h}{x}\right)^2}-\displaystyle\frac{\dd}{\dd{x}}{\mc H}\left(h\frac{\dd}{\dd{x}}{\mc H}\left(h\frac{\dd}{\dd{x}}{\mc H}[h]\right)\right)-\frac{1}{2}\frac{\dd}{\dd{x}}{\mc H}\bigg[h^2\frac{\dd^2{h}}{\dd{x}^2}\bigg]\right\}\\
&\phantom{=\,}&+ {\mc O}(h^4).
\end{eqnarray*}
that is, the formula (\ref{evap_expn}).
\section{Hilbert transform}
\label{hbt}
\subsection{Definition and basic properties}
Given a bounded function $f(x)$, we define the Hilbert transform with the usual conventions:
\[
    {\mc H}[f](x)=\frac{1}{\pi}\lim_{\varepsilon\to0^+}
\int_{|x-x'|>\epsilon}\dd{x'}\frac{f(x')}{x'-x}
\]
where we have taken the Cauchy principal value (symmetric limit) at $x'=x$. Useful properties are commutation with linear differential operators and the inversion relation ${\mc H}^{-1}=-{\mc H}$. With this definition, the Hilbert transform is not defined for constant functions. However, one can remove the divergence at infinity by taking the principal values both at $x$ and at infinity; the result is then ${\mc H}[\rmn{Cst}]=0$. (Note that the inversion formula is not valid for constants.)
\subsection{Hilbert transform and slow space varying amplitude}
In the weakly non-linear analysis, we have to compute quantities of the form ${\mc H}[A(\epsilon x)\ed^{ikx}]$, with $0<\epsilon\ll1$. We want here to show that, up to a very good precision (for $\epsilon$ sufficiently small), we have the relation 
\begin{equation}
\label{permut}
{\mc H}[A(\epsilon x)\ed^{ikx}]=A(\epsilon x){\mc H}[\ed^{ikx}] \qquad(k\neq0)
\end{equation}
which means that the action of the Hilbert transform on a slowly modulated Fourier mode does not introduce, except if $k=0$, non-localities \cite[see][]{engel}. In the next appendix, we make a substantial use of this property.

Since $A_\epsilon(x)=A(\epsilon x)$ varies significantly only if $x$ has a variation of order $1/\epsilon$, the Fourier transform $\hat{A}_\epsilon$ of $A_\epsilon$ must be negligible outside of $(-\epsilon,\epsilon)$. 

Let's first assume that the support of $\hat{A}_\epsilon$ (that is, the domain where it has non zero values) is included in $(-\epsilon,\epsilon)$. Then, using ${\mc H}[\ed^{ikx}]=i\,\rmn{sgn}(k)\ed^{ikx}$ ($\rmn{sgn}(k)=\pm1$ if $\pm k>0$), we have 
\[
{\mc H}[A_\epsilon(x)\ed^{ikx}]=i\ed^{ikx}\int^\epsilon_{-\epsilon}\dd{k'}\,\hat{A}_\epsilon(k')\,\rmn{sgn}(k'+k)\ed^{ik'x}.
\]
Thus, (\ref{permut}) is proofed  if we suppose $|k|>\epsilon$, that is, if $\epsilon$ is small enough.

If $\hat{A}_\epsilon=0$ does not vanish outside of $(-\epsilon,\epsilon)$, one can show that, under the same hypothesis on $k$, to the first order in $\epsilon$, the error on (\ref{permut}) is of the order of $\int^\infty_{1/\epsilon}|\hat{A}(k)|\dd{k}$.

\section{Weakly non linear analysis}
\label{wnl}
Here we detail the weakly non linear analysis leading to the amplitude equation (\ref{amplt}). We consider the neighbourhood of the stability limit, so that we rescale the control parameter according to
\begin{equation}
\Omega(k_\rmn{c})=\epsilon^2\omega(k_\rmn{c}).
\label{scale_omega}
\end{equation}
Plugging (\ref{evap_expn}) into the lubrication equation (\ref{lubr2}), we obtain a closed integro-differential equation for the height profile $h$. We assume that $h$ is a function of $x,X,t,T$ (fast and slow scales) and admits the expansion $h=\epsilon h^{(1)}+\epsilon^2h^{(2)}+\cdots$. The derivatives are substituted according to
\begin{equation}
\label{fast_slow}
\partial_x\to\partial_x+\epsilon\partial_X,\quad\partial_t\to\partial_t+\epsilon^2\partial_T.
\end{equation}

It is convenient to separate the evolution operator into its linear contributions ${\mc L}$ and non-linear ones ${\mc N}$. In particular, applying transformations (\ref{fast_slow}) leads to the expansion ${\mc L}={\mc L}_\rmn{c}+\epsilon {\mc L}^{(1)}+\epsilon^2{\mc L}^{(2)}+\epsilon^3{\mc L}^{(3)}+\epsilon^4{\mc L}^{(4)}$ with
\begin{eqnarray*}
  {\mc L}_\rmn{c}&=&\displaystyle\frac{1}{3\ca}\partial_x^4+\frac{\ma}{2\ca}(\partial_{x}^2-\partial_{x}^3{\mc H})-\partial_x{\mc H},\\
{\mc L}^{(1)}&=&\left\{\displaystyle\frac{4}{3\ca}\partial_{x}^3+\frac{\ma}{2\ca}(2\partial_x-3\partial_{x}^2{\mc H})-{\mc H}\right\}\partial_X,\\
{\mc L}^{(2)}&=&\left\{\displaystyle\frac{2}{\ca}\partial_{x}^2+\frac{\ma}{2\ca}(1-3\partial_x{\mc H})\right\}\partial_X^2-\omega(k_\rmn{c})+\partial_T,\\
{\mc L}^{(3)}&=&\left\{\displaystyle\frac{4}{3\ca}\partial_x-\frac{\ma}{2\ca}{\mc H}\right\}\partial_X^3,\\
{\mc L}^{(4)}&=&\displaystyle\frac{1}{3\ca}\partial_X^4.
\end{eqnarray*}
Note that the null-space of ${\mc L}_\rmn{c}$ contains slow space varying height profiles (i.e. functions of $X$). We now proceed to the solution order by order.
\bigskip
\begin{flushleft}
{\boldit Order $\epsilon^1$:}
\end{flushleft}

We have simply:
\begin{equation}
\label{ord1}
{\mc L}_\rmn{c}h^{(1)}=0.
\end{equation}
Using (\ref{permut}), the solution is
\[
h^{(1)}=(A_{11}(X,T)\ed^{i\kc x}+\cc)+A_{10}(X,T)
\]
$\kc$ being the critical wavenumber given by the linear stability analysis.
\bigskip
\begin{flushleft}
{\boldit Order $\epsilon^2$:}
\end{flushleft}

The equation has the form
\begin{equation}
\label{ord2}
{\mc L}_\rmn{c}h^{(2)}=-{\mc L}^{(1)}h^{(1)}-{\mc N}^{(2)}(h^{(1)})
\end{equation}
The non-linear term contains a $k=0$ mode. As the right-hand side of (\ref{ord2}) must be orthogonal to the null space of ${\mc L}_\rmn{c}$, it implies $h^{(1)}=0$. Thus, ${\mc L}_\rmn{c}h^{(2)}=0$, hence $h^{(2)}=(A_{21}(X,T)\ed^{i\kc x}+\cc)+A_{20}(X,T)$.
\bigskip
\begin{flushleft}
{\boldit Order $\epsilon^3$:}
\end{flushleft}

${\mc N}^{(3)}=0$ as $h^{(1)}=0$:
\begin{equation}
\label{ord3}
{\mc L}_\rmn{c}h^{(3)}=-{\mc L}^{(1)}h^{(2)}.
\end{equation}
Again, right-hand side of (\ref{ord3}) has to be orthogonal  to the null space of ${\mc L}_\rmn{c}$, so we have $A_{20}=0$. Hence, $h^{(3)}=(A_{31}(X,T)\ed^{i\kc x}+\cc)+A_{30}(X,T)$.
\begin{flushleft}
{\boldit Order $\epsilon^4$:}
\end{flushleft}
\begin{equation}
\label{ord4}
{\mc L}_\rmn{c}h^{(4)}=-{\mc L}^{(2)}h^{(2)}-{\mc L}^{(1)}h^{(3)}-{\mc N}^{(4)}(h^{(2)}).
\end{equation}
The non linear term is:
\[
{\mc N}^{(4)}(h^{(2)})=(f_4A_{21}^2\ed^{2i\kc x}+\cc)-\kc^2A_{21}\overline{A_{21}}
\]
with ($\overline{A_{21}}$ is the complex conjugate of $A_{21}$)
\begin{eqnarray*}
  f_4(\kc)&=&\displaystyle\frac{\ma}{\ca}\left(-2\kc^2-\kc^4-4\kc^3\right)+\frac{2\kc^4}{\ca}+\frac{1}{2}\kc^2.
\end{eqnarray*}
So, introducing $\alpha=\left(-\frac{2}{\ca}\kc^2+\frac{\ma}{2\ca}(1+3\kc)\right)$, we have from (\ref{ord4})
\begin{equation}
  -\partial_X{\mc H}[A_{30}]=\kc^2|A_{21}|^2,
\label{a30}
\end{equation}
\begin{equation}
  -\omega(\kc)A_{21}+\alpha\partial_X^2A_{21}+\partial_TA_{21}=0.
\label{a21}
\end{equation}
The solution to (\ref{ord4}) is
\[
  h_4=(A_{42}(X,T)\ed^{2ik_{\rm c}x}+\,{\rm c.c})+(A_{41}(X,T)\ed^{ik_{\rm c}x}+\,{\rm c.c})+A_{40}(X,T)
\]
with $A_{42}=\frac{f_4(\kc)}{\Omega(2\kc)}A_{21}^2$.
\bigskip
\begin{flushleft}
{\boldit Order $\epsilon^5$:}
\end{flushleft}
\begin{equation}
\label{ord5}
{\mc L}_\rmn{c}h^{(5)}=-{\mc L}^{(3)}h^{(2)}-{\mc L}^{(2)}h^{(3)}-{\mc L}^{(1)}h^{(4)}-{\mc N}^{(5)}(h^{(2)},h^{(3)}).
\end{equation}
It is convenient to decompose the non-linear term:
\[
{\mc N}^{(5)}(h^{(2)})=({\mc N}_{52}\ed^{2i\kc x}+\cc)+({\mc N}_{51}\ed^{i\kc x}+\cc)+{\mc N}_{50},
\]
where
\begin{eqnarray*}
  {\mc N}_{52}&=&f_5A_{31}A_{21}+ig_5\partial_XA_{21}^2\\
{\mc N}_{51}&=&j_5A_{30}A_{21}\\
{\mc N}_{50}&=&-\kc^2A_{21}\overline{A_{31}}-i\kc A_{21}\partial_X\overline{A_{21}}-\kc{\mc H}\left[\overline{A_{21}}\partial_XA_{21}\right]+\cc
\end{eqnarray*}
and with
\begin{eqnarray*}
  f_5(\kc)&=&\frac{\ma}{\ca}\left(-8\kc^3-2\kc^4-4\kc^2\right)+\frac{4\kc^4}{\ca}+\kc^2\\
g_5(\kc)&=&\frac{\ma}{\ca}\left(2\kc+2\kc^3+6\kc^2\right)-\frac{4\kc^3}{\ca}-\frac{\kc}{2}\\
j_5(\kc)&=&\frac{\ma}{\ca}\left(-\frac{3}{2}\kc^3-\kc^2\right)+\frac{\kc^4}{\ca}.
\end{eqnarray*}
Imposing again that the restriction of the right-hand side of (\ref{ord5}) to the null space of ${\mc L}_\rmn{c}$ has to vanish and using (\ref{a30}), we get
\begin{equation}
  i\beta\partial_X^3A_{21}+\left\{\alpha\partial_X^2-\omega(\kc)+\partial_T\right\}A_{31}+j_5A_{21}\textstyle\int{\mc H}[ |A_{21}|^2]=0
\label{a31}
\end{equation}
and
\begin{equation}
  \label{a40}
  \partial_X{\mc H}[A_{40}]=(-\omega(\kc)+\partial_T)A_{30}+\frac{\ma}{2\ca}\partial_X^2A_{30}+{\mc N}_{50}.
\end{equation}
with $\beta=\frac{4}{3\ca}\kc-\frac{\ma}{2\ca}$. 

The solution at this order is
\[
  h_5=(A_{52}(X,T)\ed^{2ik_{\rm c}x}+\,{\rm c.c})+(A_{51}(X,T)\ed^{ik_{\rm c}x}+\,{\rm c.c})+A_{50}(X,T)
\]
with $A_{52}=\frac{1}{\Omega(2\kc)}\left({\mc N}_{52}+i\Omega'(2\kc)\partial_XA_{42}\right)$.

Even if (\ref{a31}) is non-linear in $A_{21}$, it does not give the nature of the bifurcation. This is why we carry on computations to next order.

\bigskip
\begin{flushleft}
{\boldit Order $\epsilon^6$:}
\end{flushleft}
\begin{equation}
\label{ord6}
{\mc L}_\rmn{c}h^{(6)}=-{\mc L}^{(4)}h^{(2)}-{\mc L}^{(3)}h^{(3)}-{\mc L}^{(2)}h^{(4)}-{\mc L}^{(1)}h^{(5)}-{\mc N}^{(6)}(h^{(2)},h^{(3)},h^{(4)}).
\end{equation}
We only need the part of ${\mc N}^{(6)}$ of wavenumber $\kc$:
\begin{eqnarray*}
  \mc{N}_{61}&=&f_{6}\overline{A_{21}}A_{42}+g_{6}A_{21}\partial_X\mc{H}A_{30}+ij_6{A_{21}}\partial_XA_{30}\\
  &\phantom{=}&\,+\ell_6({A_{21}}A_{40}+A_{31}A_{30})+{m}_6|A_{21}|^2A_{21}+in_6{A_{30}}\partial_XA_{21}
\end{eqnarray*}
with
\begin{eqnarray*}
  f_6(\kc)&=&\textstyle\frac{\ma}{\ca}\left(-\kc^2-\frac{9}{2}\kc^3\right)+\frac{7\kc^4}{\ca}\\
  g_6(\kc)&=&\textstyle\frac{\ma}{\ca}\left(\frac{1}{2}\kc^3+\frac{1}{2}\kc^2\right)-\kc\\
  j_6(\kc)&=&2\textstyle\frac{\ma}{\ca}(\kc^3+\kc^2+\kc)-\frac{\kc^3}{\ca}\\
  \ell_6(\kc)&=&-\textstyle\frac{\ma}{\ca}\left(\frac{3}{2}\kc^3+\kc^2\right)+\frac{\kc^4}{\ca}=j_5(\kc)\\
  {m}_6(\kc)&=&\textstyle\frac{\ma}{\ca}\left(\frac{7}{4}\kc^5-\frac{1}{2}\kc^2-\frac{3}{4}\kc^4-\frac{5}{2}\kc^3\right)-\frac{7\kc^3}{2}+\frac{\kc^4}{\ca}\\
  n_6(\kc)&=&\textstyle\frac{\ma}{\ca}\left(\frac{9}{2}\kc^2+2\kc\right)-\frac{4\kc^3}{\ca}
\end{eqnarray*}
After solving equations ({\ref{a30},\ref{a40}}) for $A_{30}$ and $A_{40}$, we use the solvability condition that the right-hand side of (\ref{ord6}) is orthogonal to the null space of ${\mc L}_c$ and we obtain an equation for $A_{41}$. Introducing
\[ A=\epsilon A_{11}+\epsilon^2 A_{21}+\epsilon^3 A_{31}+\epsilon^4 A_{41}+\cdots, \]
the last equation for $A_{41}$ can be re-summed with equations (\ref{a21},\ref{a31}) for $A_{21}$ and $A_{31}$. Using the inverse transformations of (\ref{scale_omega},\ref{fast_slow}), we finally obtain the amplitude equation for $A(X,T)$:
\begin{eqnarray*}
(-\Omega(\kc)+\alpha\partial_X^2+\partial_T)A+i\beta\partial_X^3A+\frac{1}{3\ca}\partial_X^4A+\xi|A|^2A&\\
+i\kc(j_6\kc-\ell_6) \mc{H}[ |A|^2]{A}+in_6\kc^2(\partial_XA)\int\mc{H}[ |A|^2] &\\
+\ell_6\bigg\{\kc^2 A\int\!\!\!\int(-\Omega(\kc)+\partial_T+\partial_X\mc{H})|A|^2+2i\kc{A}\int\mc{H}\left[A\partial_X\overline{A}\right]\bigg\}&=0.
\end{eqnarray*}
with
\[
  \xi(\kc)={m}_6+\frac{1}{\Omega(2\kc)}f_6f_{22}-\kc^2g_6+\left(\kc^2\frac{\ma}{2\ca}-2\kc\right)\ell_6.
\]

Taking the limit of small $\kc$ with the help of the rescaling  $\partial_X\to\kc\partial_X$ and $\partial_T\to\kc\partial_T$ leads to
\begin{equation}
  \label{non_loc}
  \begin{array}{r}
    \medskip  \displaystyle\left(\frac{\partial}{\partial T}-\sigma-\frac{3}{2}\frac{\partial^2}{\partial X^2}\right)A+2i\frac{\partial^3A}{\partial X^3}+\frac{1}{2}\frac{\partial^4A}{\partial X^4}+\frac{45}{8}|A|^2A+i6\kc \mc{H}[ |A|^2]{A}\,+\phantom{=0}\\
    -\displaystyle\frac{3}{2}\kc\left\{ A\int\!\!\!\int\left(\frac{\partial}{\partial T}-\sigma+\frac{\partial}{\partial X}\mathcal{H}\right)|A|^2+2i{A}\int\mc{H}\left[A\frac{\partial\overline{A}}{\partial X}\right]\right\}+{\mc O}(\kc^2)=0,
  \end{array}
\end{equation}
hence the simplification (\ref{amplt}). Note that (\ref{non_loc}) is valid only if $A(X,T)$ vanishes at $X=\pm\infty$, so that Hilbert transforms ${\mc H}$ are well defined. In contrast with standard weakly non-linear analysis, this equation is non-local. 


\bibliography{evapo}
\bibliographystyle{jfm}
\end{document}